\documentclass[12pt]{iopart}
\usepackage{iopams}
\expandafter\let\csname equation*\endcsname\relax
\expandafter\let\csname endequation*\endcsname\relax
\usepackage{amsmath,amssymb,bm,yfonts}
\usepackage[pdftex]{hyperref}
\usepackage{cite}
\usepackage[pdftex]{graphicx}
\usepackage[utf8]{inputenc}
\newcommand{\D}{\nabla}
\newcommand{\ts}{\textstyle}
\newcommand{\M}{\mathcal M}

\makeatletter
\g@addto@macro\@floatboxreset\centering
\makeatother

\begin{document}

\title[Singularities in rotating black holes coupled to a massless scalar field]
{Singularities in rotating black holes  coupled to a massless scalar field}


\author{Paul~M.~Chesler}
\address{Black Hole Initiative, Harvard University, Cambridge, MA 02138, USA}
\ead{pchesler@g.harvard.edu}
\vspace{10pt}

\begin{indented}
\item[] \today
\end{indented}

\begin{abstract}
	We employ a late-time expansion to study the interior of rotating black holes coupled to a massless scalar field
	in asymptotically flat spacetime.  We find that
	decaying fluxes of scalar radiation into the black hole necessitate the existence of 
	a null singularity at the Cauchy horizon and a central spacelike singularity at radius $r = 0$. 
    In particular, 
	the decaying influxes source a localized cloud of scalar radiation near $r=0$ whose amplitude grows unboundedly large as advanced time $v \to \infty$. The scalar cloud inevitably results in a central spacelike singularity at late times $v$, with the curvature near $r = 0$ diverging like $r^{-\alpha v}$, where $\alpha > 0$ is a constant.
%
%
\end{abstract}

%
%
%
%
%

\section{Introduction}
Long after black holes form, their exterior geometry should relax to the Kerr-Newman geometry.  Perturbations relax by either being absorbed by the horizon or by being radiated to infinity.  However, deep inside the black hole no such relaxation mechanism exists.  This  means that the interior geometry is not unique and depends on initial conditions.

\begin{figure}[h]
	\includegraphics[trim= 0 0 0 0 ,clip,scale=0.4]{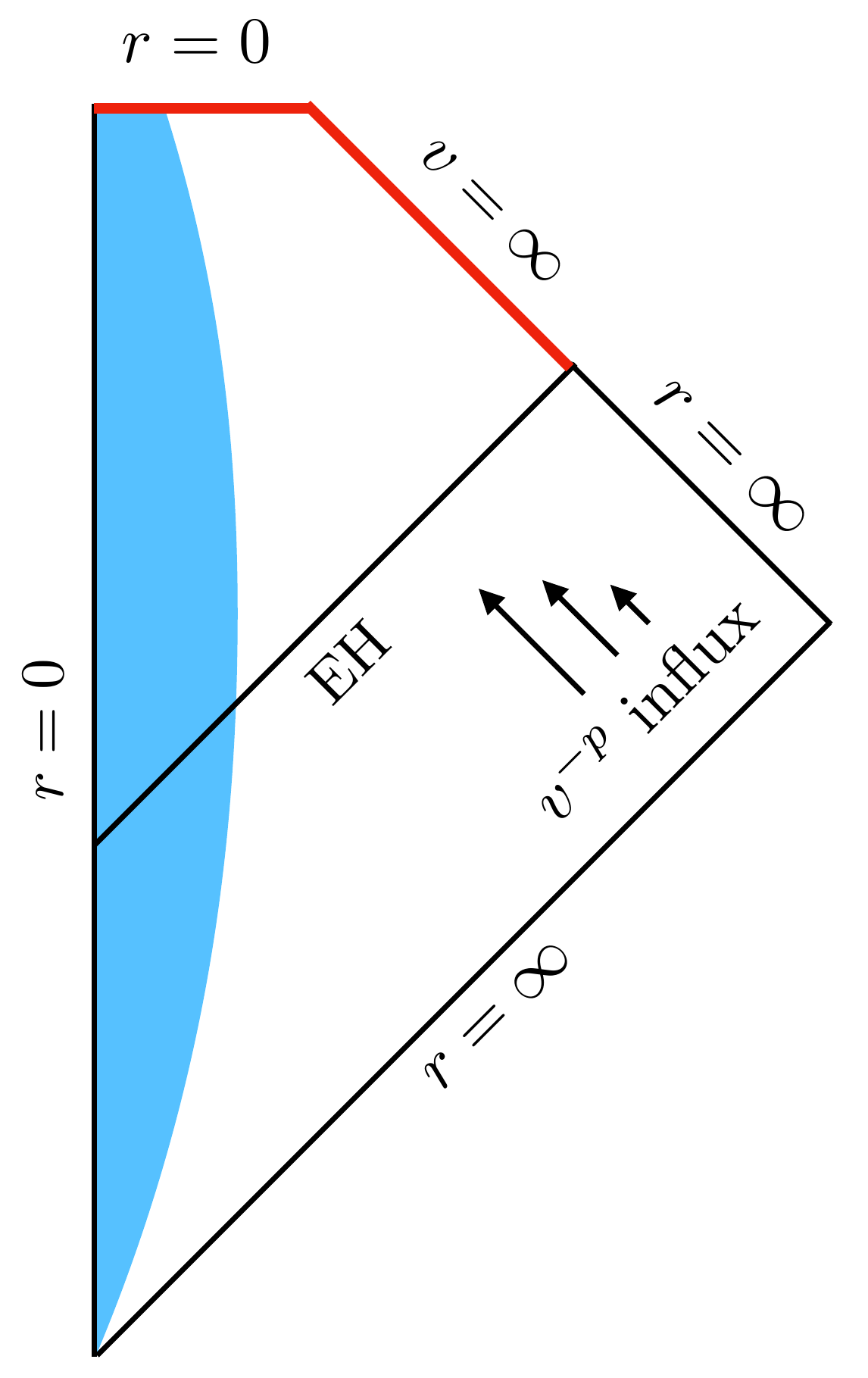}
	\caption{
		A Penrose diagram showing gravitational collapse of matter (blue shaded region).  During collapse an event horizon (EH) forms. Outgoing radiation emitted during collapse scatters off the gravitational potential at large distances, resulting in an influx of radiation which decays like $v^{-p}$.  The geometry contains a null singularity at $v = \infty$ and a 
		spacelike singularity at $r = 0$ (both shown as red lines).
	}
	\label{fig:PenroseDiagram}
\end{figure}

There nevertheless do exists universal features in the interior of black holes such as singularities \cite{Penrose:1964wq}.  
One type of singularity is a null singularity on the geometry's Cauchy Horizon (CH), located at  advanced time $v = \infty$.  Consider gravitational collapse in asymptotically flat spacetime, as  depicted in the Penrose diagram in Fig.~\ref{fig:PenroseDiagram}.
Energy radiated outwards during collapse scatters off the gravitational potential at large distances, resulting in decaying influxes of radiation through the event horizon (EH).  Price reasoned the influx through the horizon decays with an inverse power of $v$ \cite{Price:1971fb,Price:1972pw}.  From the perspective of  observers crossing the CH, this influx appears blue shifted by a factor of $e^{\kappa v}$, with $\kappa$ the surface gravity of the inner horizon. Penrose reasoned this 
exponential blue shift leads to a curvature singularity on the CH \cite{Penrose:1968ar}.  Numerous studies \cite{Poisson:1989zz,PhysRevD.41.1796,Simpson:1973ua,HISCOCK1981110,PhysRevD.20.1260,PhysRevLett.67.789,0264-9381-10-6-006,Brady:1995ni,Burko:1997zy,Hod:1998gy,Burko:1997fc,10.2307/3597235,Dafermos:2017dbw,Ori:2001pc,Ori1997,Burko:2016uvr,Kaplan:2018dqx,CNC} demonstrate that a null singularity at the CH is a generic feature of black holes in asymptotically flat spacetime 
\footnote{A notable exception to this is near-extremal black holes in de Sitter spacetime \cite{Cardoso:2017soq}.}.

In this paper we shall primarily focus on central singularities in black holes created via collapse.
Numerical simulations of the collapse of a spherically symmetric charged
massless scalar field are consistent with the formation of a central spacelike singularity at areal radius $r = 0$  \cite{Hod:1998gy} (see also \cite{Brady:1995ni,Burko:1997zy}).  Combined with the singular CH, this means the singular structure of the spacetime is that shown in the Penrose diagram in Fig.~\ref{fig:PenroseDiagram} \cite{Hod:1998gy}. How generic are these results? 
In \cite{CNC} we argued that decaying influxes of radiation produced in spherically symmetric collapse make the formation of a spacelike singularity inevitable.  The 
infalling radiation forms a localized cloud of scalar radiation near $r = 0$, whose amplitude grows unboundedly large as $v \to \infty$, and which inevitably 
results in the formation of a spacelike singularity.  As a consequence of the growing cloud amplitude, 
the strength of the singularity increases with time, with the
curvature  diverging like $r^{-\alpha v}$ where $\alpha > 0$ is a constant.
This behavior is universal, as it arises from the universal power law influx of radiation into the black hole. Our goal in this paper is to generalize the analysis of \cite{CNC} to rotating black holes.  

It was argued long ago by Belinsky, Khalatnikov, and Lifshitz that as a generic spacelike singularity is approached, the geometry oscillates wildly \cite{Belinsky:1970ew,doi:10.1063/1.3382327}.  It was subsequently 
pointed out that the oscillatory structure can be ameliorated  
with the introduction of a scalar field \cite{Belinski:1973zz}.
With this simplification in mind, we choose to study black holes coupled to a massless scalar field.

\begin{figure}[h]
	\includegraphics[trim= 0 0 0 0 ,clip,scale=0.35]{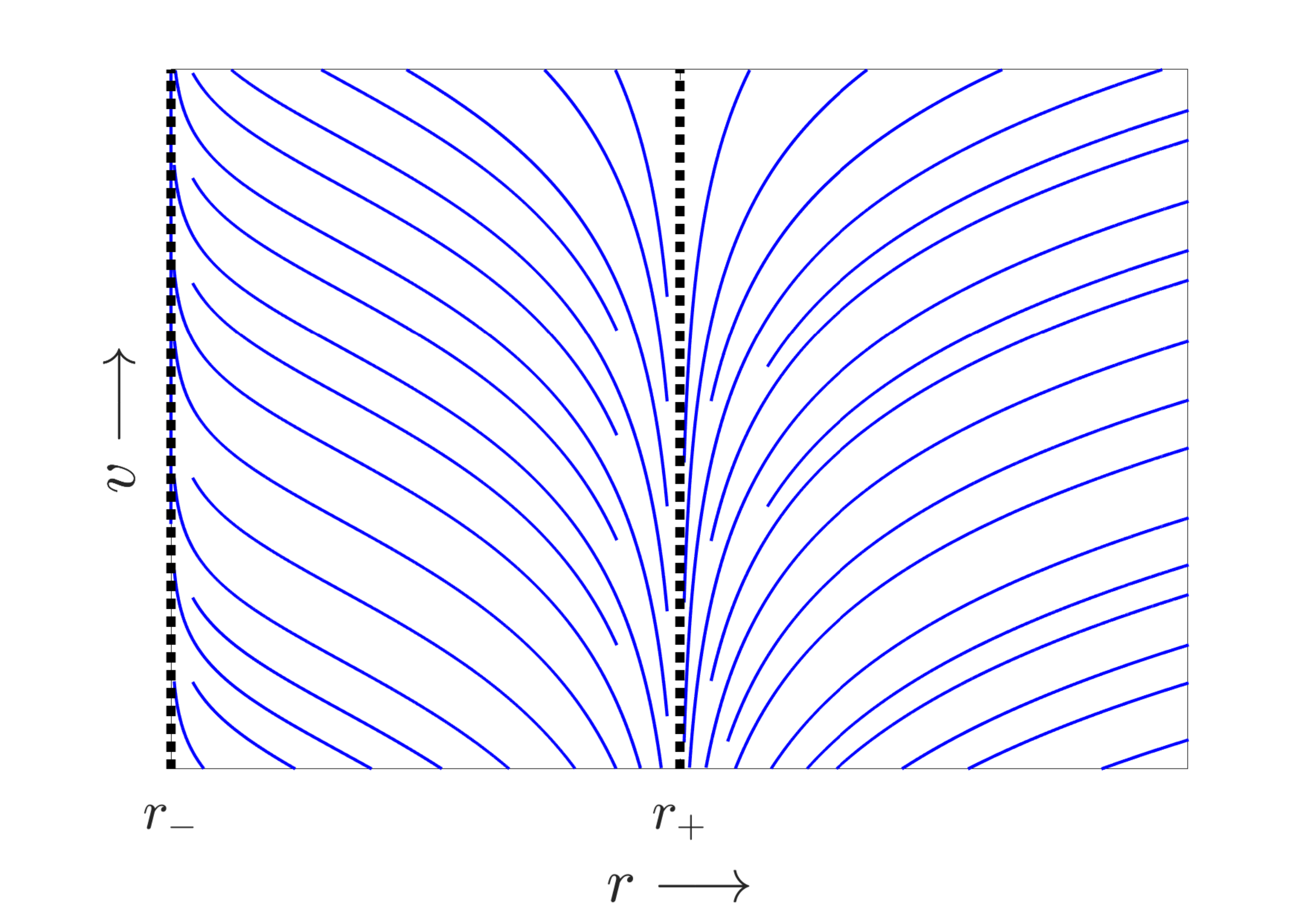}
	\caption{
    A sketch of some outgoing null geodesics in the Kerr geometry.  Geodesics at $r > r_+$ can escape to $r = \infty$.  Geodesics at $r_- < r < r_+$ asymptote to $r = r_-$ as $v \to \infty$.  Note radial infalling null geodesics are lines of constant advanced time $v$.
	}
	\label{fig:OutgoingGeos}
\end{figure}

Our analysis closely mirrors that of Ref.~\cite{CNC} and is based on two key assumptions.  First, we assume there is an influx of scalar radiation through the horizon which decays like $v^{-p}$.
We assume $p$ is sufficiently large such that the mass $M(v)$ and spin $a(v)$ of the black hole approach constants $M(\infty)$ and $a(\infty)$
as $v \to \infty$.   
The infalling radiation 
can scatter off the gravitational field, thereby exciting outgoing radiation in the interior of the black hole.  This, together with outgoing radiation emitted during collapse, means that the interior of the black hole is filled with outgoing radiation.

Recall that the Kerr solution contains inner and outer horizons at radii $r_-$ and $r_+$, respectively (we define our coordinates below in Sec.~\ref{sec:eqm}).  
The radii $r_\pm = r_\pm(M,a)$ depend on the mass and spin of the black hole.  Following \cite{CNC}, our second assumption is that the geometry exterior to the would-be inner horizon relaxes to the Kerr solution as $v\to\infty$ %
\footnote{For the Kerr solution the surface $r = r_-$ is null and an inner apparent horizon. Out of equilibrium this need not be the case.	  
}.  
Why is this a reasonable assumption?  In the Kerr geometry, all light rays inside the black hole propagate to $r \leq r_-$ as $v \to \infty$. 
See Fig.~\ref{fig:OutgoingGeos} for a sketch of some outgoing light rays in the $v{-}r$ plane.
Therefore, the Kerr geometry naturally provides a mechanism for perturbations at $r > r_-$ to relax; perturbations naturally propagate to $r \leq r_-$. Numerical simulations of the interior of rotating black 
holes are consistent with the geometry at $r > r_-$ relaxing to the Kerr solution as $v \to \infty$ \cite{Chesler:2018hgn}.

The assumption that the geometry exterior to $r_-$ relaxes to the Kerr solution has dramatic consequences on infalling observers passing through $r_-$.  As sketched in Fig.~\ref{fig:OutgoingGeos}, in the Kerr geometry outgoing light rays between $r_- < r < r_+$ approach $r_-$ as $v \to \infty$.  This means that outgoing radiation inside the black hole must be localized to a ball whose surface approaches $r_-$ as $v\to \infty$.  From the perspective of infalling observers, the outgoing radiation appears blue shifted by a factor of $e^{\kappa v}$.
In follows that infalling observers encounter an effective ``shock" at $r = r_-$ \cite{Marolf:2011dj,Eilon:2016osg,Chesler:2018hgn,Burko:2019fgt}, 
where the blue shifted radiation results in a Riemann tensor jumping from its Kerr limit to order $e^{2 \kappa v}$.  
Correspondingly, Raychaudhuri's equation implies the ball of blueshifted  
radiation focuses infalling radial geodesics.  The focusing means that radial infalling time-like observers propagate from $r = r_-$ to $r = 0$ over a proper time $\Delta \tau \sim e^{-\kappa v}$ \cite{Marolf:2011dj}.  In other words, infalling time-like observers encounter $r = 0$ effectively instantaneously after passing through the shock at $r = r_-$.
Likewise, null rays are focused from $r = r_-$ to $r = 0$ over an affine parameter interval
\begin{equation}
\label{eq:Dlambda}
\Delta \lambda \sim e^{-\kappa v}.
\end{equation}
The scaling (\ref{eq:Dlambda}) has been verified numerically for non-equilibrium black holes in \cite{Eilon:2016osg,Chesler:2018hgn}.

\begin{figure}[h]
	\includegraphics[trim= 0 0 0 0 ,clip,scale=0.6]{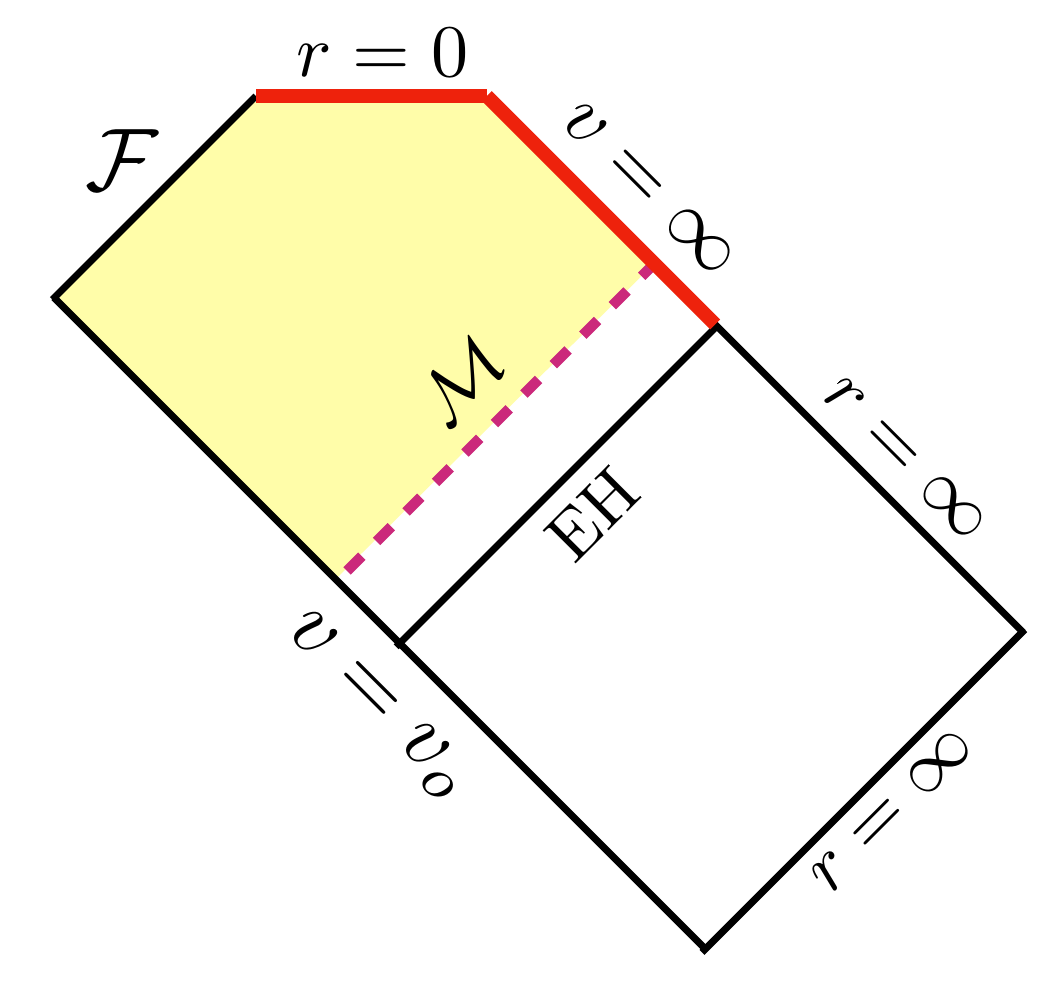}
	\caption{
		A Penrose diagram showing the initial value problem we study in this paper. Initial data is specified on some infalling $v = v_o$ null surface, with $v_o$ taken arbitrarily large. Boundary data is specified on some outgoing null surface $\M$.
		In the shaded region we solve the equations of motion with a derivative expansion in $\lambda$.  The outgoing null surface $\mathcal F$ bounds the inner domain of validity 
		of the derivative expansion initial value problem and intersects $r = 0$ at finite time $v$.
		Decaying $v^{-p}$ fluxes of infalling radiation through $\M$ always results in a null singularity at $v = \infty$
		and a spacelike singularity at $r = 0$, irrespective of initial data at $v = v_o$.
	}
	\label{fig:CompDomain}
\end{figure}

As was first pointed out in \cite{CNC}, the focusing of infalling geodesics by shocks suggests  there
exists an expansion parameter $\epsilon \equiv e^{-\kappa v} \ll 1$ 
in terms of which the equations of motion can be solved perturbatively in the region $r<r_-$ at late times.  This can
be made explicit by employing the affine parameter $\lambda$ of infalling null geodesics as a radial coordinate and expanding the equations of motion inside $r_-$ in a derivative expansion in $\lambda$.  
Such an expansion is simply an expansion in powers of $\epsilon$, which is arbitrarily small at late times.
We shall see that at leading order in the derivative expansion, Einstein's equations reduce to a set of 1+1D PDEs at each set of angles.  This 
reflects the fact that as $v \to \infty$, dynamics at each set of angles become causally disconnected.  

The yellow shaded region in Fig.~\ref{fig:CompDomain} shows the domain of dependence of the initial value problem we consider in this paper. Initial data is specified on some infalling $v = v_o$ null surface and boundary data is specified on some outgoing null surface $\M$ inside the EH.  We consider the limit where $v_o$ is arbitrarily large.  On $\M$ we impose the boundary condition that there is an influx of scalar radiation decaying like $v^{-p}$.  The outgoing null surface $\mathcal F$ bounds the inner domain of validity of the derivative expansion initial value problem and intersects $r = 0$ at a finite $v$.  

We find that the singular structure of rotating black holes is identical
to that of spherically symmetric charged black holes \cite{CNC}.  In particular, infalling radiation necessitates the existence of a null singularity on the CH and results in the formation of a localized cloud of scalar radiation near $r = 0$ whose amplitude grows unboundedly large as $v \to \infty$. 
The growing cloud always results in a spacelike singularity at $r = 0$ at late enough times, 
irrespective of initial conditions at $v = v_o$, with the curvature 
diverging like $r^{-\alpha v}$ with $\alpha > 0$.

An outline of the rest of the paper is as follows.  In Sec.~\ref{sec:eqm} we
write the equations of motion, employing the affine parameter $\lambda$ as a radial coordinate. In Sec.~\ref{sec:derexpansion} we derive approximate equations of motion, valid in the limit of large $\lambda$ derivatives.  In Sec.~\ref{sec:singularstruct} we employ the approximate equations of motion to study the causal structure of singularities inside black holes, and in Sec.~\ref{sec:discussion} we discuss generalizations of our work.

\section{Equations of motion}
\label{sec:eqm}
Consider Einstein's equations coupled to a real massless scalar field $\Psi$.
The equations of motion read
\begin{align}
\label{eq:eqmKerr}
&E_{\mu \nu} \equiv R_{\mu \nu} + {\textstyle \frac{1}{2}} R g_{\mu \nu} - 8 \pi T_{\mu \nu} = 0, &
& D^2 \Psi = 0,&
\end{align}
where $D_\mu$ is the covariant derivative and
\begin{subequations}
	\label{eq:stressKerr}
	\begin{align}
	T_{\mu \nu} &= D_\mu \Psi  D_\nu \Psi -\frac{1}{2} g_{\mu \nu} ( D \Psi)^2, 
	\end{align}
\end{subequations}
is the scalar stress
tensor.

Following \cite{CNC,Chesler:2013lia}, we employ a characteristic evolution scheme where the metric takes the form
\begin{equation}
\label{eq:metricKerr}
ds^2 =  -2 A  dv^2 + 2  d\lambda dv 
+ G_{ab} (dx^a - F^a dv)(dx^b - F^b dv).
\end{equation}
Here $x^a = \{\theta,\phi\}$ are polar and azimuthal angular coordinates, $\lambda$ is the radial coordinate, and $v$ is the time coordinate.   Here and in what follows, lower case latin indices $a,b,c,\dots,$ run over the two angular indices.  It follows from the geodesic equations that radial infalling null geodesics satisfy by $v = {\rm const.}$, $x^a = {\rm const.}$, and $d^2 \lambda/ds^2 = 0$, where $s$ is an affine parameter.  This means  $\lambda$ itself affinely parameterizes null infalling radial geodesics.
The metric (\ref{eq:metricKerr}) is invariant under $\lambda$-independent coordinate changes of $x^a$,
\begin{align}
\label{eq:2ddiffs}
x^a \to \bar x^a(v,x^b),
\end{align}
and the  $\lambda$-independent shifts
\begin{equation}
\label{eq:resdiffKerr}
\lambda \to \lambda + \xi(v,x^a),
\end{equation}
for arbitrary $\xi$.

We will write Einstein's equations in component form below, keeping the angular dependence covariant.  In doing so we define $\D$ to be the covariant derivative operator w.r.t$.$  the 2D angular metric $G_{ab}$.  We shall raise and lower angular indices with $G_{ab}$ and $G^{ab} \equiv (G^{-1})_{ab}$. In addition to $\D$, we will write Einstein's equations in terms of the 
the derivative operators
\begin{align}
&' \equiv \partial_\lambda, &
d_+ \equiv \partial_v + A \partial_\lambda.&
\end{align}
The $'$ derivative is just the directional derivative along infalling radial null geodesics whereas
$d_+$ is the directional derivative along an outgoing null geodesics.
These operators don't commute with raising and lowering  angular indices. We adopt the convention  that $d_+$ and $'$ derivatives of all 2D vector and tensor quantities will be understood to be taken with all indices down before raising any indices.  For example, $F'^{a} \equiv G^{ab} \partial_\lambda F_{b}$ and $d_+ F^{a} \equiv G^{ab} (\partial_v  + A \partial_\lambda) F_{b}$. Lastly, we define the normalized angular metric $h_{ab}$ and areal coordinate $r$ via
\begin{equation}
G_{ab} = r^2 h_{ab},
\end{equation}
where $h_{ab}$ satisfies 
\begin{equation}
\label{eq:hnorm}
\det h_{ab} = \sin^2 \theta.
\end{equation}

With the above conventions and definitions, we can now write Einstein's equations in component form. As we shall see, requisite initial data at $v= v_o$ consists of $h_{ab}$ and $\Psi$.  The Einstein-scalar system decomposes into a set of dynamical equations and a set of radial constraint equations.  The dynamical equations take of the form of a nested system of linear ODEs in $\lambda$ 
for $\partial_v h_{ab}$ and $\partial_v \Psi$ \cite{Chesler:2013lia}.  The radial constraint equations need only be enforced on some $\lambda = {\rm const.}$ surface and can be imposed as boundary conditions on $\M$.  We begin with the dynamical set of equations.

The component $E^{vv} = 0$ of Einsteins equations yields 
\begin{equation}
\label{eq:req}
r'' + \left [ {\textstyle \frac{1}{8}} \tr (h^{-1} h' h^{-1} h')   + 4 \pi \Psi'^2\right] r = 0,
\end{equation}
Hence if $h_{ab}$ and  $\Psi$ are specified, Eq.~(\ref{eq:req}) can be integrated in from $\M$ to find $r$.  It turns out that Eq.~(\ref{eq:req})
is just Raychaudhuri's equation for infalling radial null geodesics.  
Note that the quantity in the brackets is $\geq 0$.  It follows that $r'$ can only increase
as $\lambda$, or equivalently $r$, decreases.  

Defining
\begin{equation}
\label{eq:Pidef}
\Pi_a = r^2 G_{ab} \partial_\lambda F^b, 
\end{equation}
the equations $E^{v}_{\ a} = 0$ yield
\begin{equation}
\label{eq:Pieq}
\Pi'_a = S_{a}^{(\Pi)},
\end{equation}
where the source function $S_{a}^{(\Pi)}$ is
\begin{equation}
S_{a}^{(\Pi)} =-16 \pi r^2 \Psi' \D_a \Psi + r^2 \D^b G'_{a b} - 4 r^2 (1/r \D_a r)'.
\end{equation}
Hence, if $h_{ab}$, $r$ and $\Psi$ are given, Eq.~(\ref{eq:Pieq}) can be integrated in from $\M$ to
find $\Pi_a$.  Likewise, with $\Pi_a$ known, Eq.~(\ref{eq:Pidef}) can be integrated to find $F^a$.

The combination $E^{v \lambda} - A E^{vv} = 0$ yields
\begin{equation}
\label{eq:rdoteq}
(r d_+ r)' = S^{(d_+r)},
\end{equation}
where the source function $S^{(d_+r)}$ reads
\begin{equation}
S^{(d_+r)} = \ts {\textstyle \frac{r^2}{4} } \mathcal R - \frac{1}{8 r^2} \Pi^2 - \frac{1}{4} \D \cdot \Pi+
\frac{1}{2r} \Pi \cdot \D r -  \left(r \D \cdot F - F\cdot \D r \right) r' - r F \cdot \D r'- 2 \pi r^2 (\D \Psi)^2,
\end{equation}
with $\mathcal R$ the 2D Ricci scalar associated with the angular metric $G_{ab}$. Hence given $h_{ab}$, $\Psi$, $r$, $F^a$ and $\Pi_a$, this equation can be integrated in from $\M$ to find $d_+ r$.

The equation $E_{ab} - \frac{1}{2} G_{ab} G^{cd}E_{cd} = 0$ yields
\begin{equation}
\label{eq:hdoteq}
\ts \left \{ \delta^{c}_{a} \delta^{d}_b - \frac{1}{2}G^{cd} G_{ab}\right \} [ (r d_+ h_{cd})' - S_{cd}^{(d_+ h)} ] = 0,
\end{equation}
where
\begin{eqnarray}
\nonumber
S_{ab}^{(d_+h)} =  &-& \ts \frac{1}{r^3} \D_a \Pi_b - \frac{1}{2 r^5} \Pi_a \Pi_b + \frac{2}{r^4} \Pi_a \D_b r
- \frac{2 r'}{r^2} \D_a F_b - \frac{1}{r} F \cdot \D G'_{ab} 
\\ &-& \ts r \D_{[a} F_{c]} G^{c d}h'_{db}
- \ts (d_+ r + \frac{1}{2} r \D \cdot F) h'_{ab} - \frac{8 \pi}{r} \D_{a}\Psi \D_b \Psi.
\end{eqnarray}
Hence, given $h_{ab}$, $\Psi$, $r$, $\Pi_a$, $F_a$ and $d_+ r$, this equation can be integrated in from $\M$ to find $d_+ h_{ab}$.

The scalar equation of motion reads
\begin{equation}
\label{eq:scalareq}
\ts (r d_+ \Psi)' = - \Psi' d_+ r - \frac{1}{2 r} \Pi^2 - r F\cdot \D \Psi' - \frac{r}{2}  \D^2 \Psi - \frac{r}{2} \D \cdot F \Psi' - r' F \cdot \D \Psi.
\end{equation}
Hence, given $h_{ab}$, $\Pi_a$, $F^a$ and $d_+ r$, this equation can be integrated in from $\M$ to find $d_+ \Psi$.

The equation $E^{a}_{\ a} - 2 E^{v}_{\ v} - F^a E^{v}_{\ a} = 0$ yields
\begin{equation}
\label{eq:Aeq}
A'' = S^{(A)},
\end{equation}
where
\begin{eqnarray}
\nonumber 
S^{(A)} =  &-& \ts \frac{r^2}{4}  h'^{ab} d_+ h_{ab}  + \frac{2 r'}{r^2} d_+ r - \frac{1}{2} \mathcal R + \frac{3}{4 r^4} \Pi^2 - \frac{1}{2}
\D^a F^b G'_{ab} + \frac{2 r'}{r} \D \cdot F \\
&-& \ts 8 \pi \Psi' d_+ \Psi + 4 \pi (\D\Psi)^2 - 8 \pi \Psi' F \cdot \D \Psi.
 \end{eqnarray}
Hence, given $h_{ab}$, $\Psi$, $r$, $\Pi_a$, $F_a$, $d_+ r$ and $d_+ h_{ab}$, this equation can be integrated in from $\M$ to find $A$.

We therefore see that given $h_{ab}$ and $\Psi$ on some $v = {\rm const.}$ slice,  and boundary data on $\M$, the Einstein-scalar scalar system reduces to a nested system of linear ODEs in $\lambda$ for $d_+ h_{ab}$, $d_+ \Psi$ and $A$.  Once
these quantities are solved for, the time derivatives of $h_{ab}$ and $\Psi$ are determined by
	\begin{align}
     &\partial_v \Psi = d_+ \Psi - A \Psi',&&
      \partial_v h_{ab} = d_+ h_{ab} - A h_{ab}'.&
	\end{align}
With the time derivatives computed, the system can then be advanced forward in time. 

The remaining components of Einstein's equations, $E^{\lambda}_{ \ v} = 0$ and $E^{\lambda}_{\ a} = 0$,
are radial constraint equations.  In particular, it follows from the Bianchi Identities that if these equations are satisfied at one value of $\lambda$ and the other components of Einstein's equations are satisfied at all values of $\lambda$, then the $E^{\lambda}_{ \ v} = 0$ and $E^{\lambda}_{\ a} = 0$ components must be satisfied at all values of $\lambda$.  Because of this, the radial constraint equations can be imposed as boundary conditions on all the other equations.  

The combination $E^{\lambda}_{ \ v} + F^a E^{\lambda}_{\ a} + A^2 E^{vv} = 0$ yields 
\begin{align}
\nonumber
\ts & \ts d_+^2 r + \frac{r}{2} \D \cdot d_+ F - \frac{r^3}{2} F^a  D^b d_+ h_{ab} + F \cdot \D d_+r - r^2 F^a D^b r d_+h_{ab} + \frac{1}{8} r^5 d_+ h^{ab} d_+ h_{ab} - \frac{1}{2} r \D^2 A \\ \nonumber
 &\ts -( 1/r F\cdot \D r +A') d_+r  - \frac{1}{2} r \D \cdot F A'-r' F \cdot \D A + \frac{1}{2} r F^a \D^2 F_a + \frac{1}{2} r D^{(a} F^{b)}D_{(a} F_{b)}
\\ \label{eq:rddoteq}
&\ts + \frac14 F^2 \mathcal R- \frac{1}{2} r F^{[a} D^{b]} D_{[a} F_{b]} - 4 \pi r d_+ \Psi^2 + 4 \pi r F^2 (\D \Psi)^2 - 4 \pi r (F^a \epsilon_{a b} \D^b \Psi)^2 = 0,
\end{align}
while the combination $E^{\lambda}_{\ a} - A E^{\lambda}_{ \ a} = 0$ yields
\begin{align}
\label{eq:Fdoteq}
(d_+ F^a)' = S^a_{(d_+ F)},
\end{align}
where
\begin{align}
\nonumber 
S^a_{(d_+F)} & =  \ts F^b (d_+ G^a_{ \ b})' - \D_b d_+ G^{a b} + \frac{4}{r} \D^a r + 2 \D^a A' - \frac{2}{r^2} F_b\left [ \D^{(a} \Pi^{b)}] - \frac{1}{2} G^{ab} \D \cdot \Pi \right ]  \\ \nonumber
& \ts - 2 \mathcal R^{ab} F_b - \frac{4 d_+r}{r^2} \D^a r - \D_b A G'^{a b}+ \frac{1}{r^2} \left (  -\frac{2}{r} d_+ r + A' - 2 \D \cdot F - \frac{2}{r} F \cdot r\right) \Pi^a  \\ \nonumber
& \ts+ \frac{1}{r^2} \epsilon^{ab} \epsilon^{cd} F_c \D_d \Pi_b - \frac{2}{r^2} \epsilon^{ab} \epsilon^{cd} \Pi_c \D_d F_b - D_b D^{[a} F^{b]} + 16 \pi d_+ \Psi \D^a \Psi + 16 \pi (\D \Psi)^2 F^a
\\
 & \ts+ 16 \pi \epsilon^{c d} F_c \D_d \Psi \epsilon^{ab} \D_b \Psi,
\end{align}
with $\epsilon_{ab} = -\epsilon_{ba}$ and $\epsilon_{\theta \phi} \equiv r^2 \sin \theta$.

\section{Derivative expansion}

\label{sec:derexpansion}

Following the arguments presented in the Introduction, in the region enclosed by $\M$ we shall solve the equations of motion in the limit where $\lambda$ derivatives are arbitrarily large. 
Our analysis will turn out to be insensitive to the precise choice of $\M$.  However, 
for pedagogical reasons it is useful to choose $\M$ to asymptote to $r = r_-$ as $v \to \infty$.  

The residual diffeomorphisms (\ref{eq:2ddiffs}) and (\ref{eq:resdiffKerr})
allows us to fix $A$ and $F^a$ on any $\lambda = {\rm \ const.}$ surface.  For convenience, at $\lambda = 0$ we set 
\begin{align}
\label{eq:AKerr}
A|_{\lambda = 0} &= 0,
\\ 
\label{eq:FMBC}
F^a|_{\lambda = 0} &= \Omega \delta^{a \phi},
\end{align}
for some $\Omega = \Omega(v)$.
It follows that curves with $\lambda = 0$ and
\begin{align}
\label{eq:kerrIHgeos}
&\frac{d \theta}{dv} = 0, &
&\frac{d \phi}{dv} = \Omega,
\end{align}
are outgoing null curves.
We choose $\M$ to be the null surface generated by these curves.  Note that choosing $\M$ to asymptote  to $r_-$ as $v \to \infty$ is tantamount to choosing initial radial data for these null curves.
Additionally, note the value of $\Omega = \Omega(v)$ is arbitrary. One could for example choose $\Omega$ to be the rotation rate of the inner horizon of the associated late time Kerr solution.  Alternatively, one could choose $\Omega = 0$, thereby working in frame co-rotating with the null curves which generate $\M$. In what follows we will leave $\Omega$ undetermined.  

In order to account for rapid $\lambda$ dependence,
we introduce a bookkeeping parameter $\epsilon$ and assume the scalings
\begin{align}
\label{eq:scalings}
&\partial_\lambda = O(1/\epsilon), && d_+ = O(\epsilon^0),
\end{align}
and study the equations of motion in the $\epsilon \to 0$ limit.
We shall see in Sec.~\ref{sec:nearM} that the $\epsilon \to 0$ limit is just the late time limit.
The scaling $d_+ = O(\epsilon^0)$ reflects the fact that 
quantities evaluated along outgoing null geodesics are not rapidly varying in $v$.  
The scaling relations (\ref{eq:scalings}) 
and the Einstein equation (\ref{eq:Pieq}) imply
\begin{equation}
\label{eq:Piscaling}
\Pi_a = O(\epsilon^0).
\end{equation}  
Likewise, scaling relations (\ref{eq:scalings}) and the Einstein equation (\ref{eq:Aeq}) imply $A'' = O(1/\epsilon)$.  Together with the boundary condition (\ref{eq:AKerr}), this means
\begin{align}
\label{eq:Ascalinghyp}
A = O(\epsilon).
\end{align}

Further boundary conditions are needed on $\M$.  Define
\begin{align}
\label{eq:Dplus}
D_+ &\equiv d_+ + F^a \partial_a, \\ \nonumber
&= \partial_v + A \partial_\lambda + F^a \partial_a,
\end{align}
which is simply the directional derivative along outgoing null curves with tangent $\ell^\mu = \{1,A,F^a\}$.  The total time derivative of any quantity $f$ on $\M$ is therefore $\frac{df}{dv} = D_+ f$.  On $\M$ we assume the influx of scalar and tensor radiation is a power law in accord with Price's Law:
\begin{subequations}
	\label{eq:PriceLaw0}
	\begin{align}
	\label{eq:scalarinflux} 
	D_+ \Psi|_{\lambda = 0} &\sim 1/v^p,
	\\
	\label{eq:tensorinflux} 
	D_+ h_{ab}|_{\lambda = 0} &\sim 1/v^q.
	\end{align}
\end{subequations}
We will assume $q > p$, meaning gravitational wave tails decay faster than scalar tails \cite{Donninger:2009tw}, but otherwise leave $p$
and $q$ arbitrary.

In addition to the Dirichlet boundary condition (\ref{eq:AKerr}), Eq.~(\ref{eq:Aeq}) also requires a Neumann boundary condition on $A$.  The requisite boundary condition can be obtain from our assumption that the geometry at $r > r_-$ relaxes to the Kerr solution as $v \to \infty$.
Note that even with this assumption, derivatives of the metric can become discontinuous across $r_-$ 
as $v \to \infty$. 
However, from the scaling $\partial_r (A') = O(\epsilon^0)$ it is reasonable to assume that 
$A'$ remains continuous across $r_-$ as $\epsilon \to 0$, or equivalently as $v \to \infty$.
This means that at late times $A'|_{\lambda = 0}$ must approach its Kerr limit,
\begin{equation}
\label{eq:kappaKerr}
A'|_{\lambda = 0} = -\kappa,
\end{equation}
with $\kappa$ the surface gravity of the associated Kerr inner horizon,
\begin{equation}
 \kappa = \frac{1}{2} \left (  \frac{1}{M(\infty) - \sqrt{M(\infty)^2 - a(\infty)^2}} - \frac{1}{M(\infty)} \right ),
\end{equation}
where again, $M(\infty)$ and $a(\infty)$ are the asymptotic  
mass and spin of the black hole.

Let us now analyze the $\epsilon \to 0$ limit of the Einstein-scalar system.  
All terms in Eqs.~(\ref{eq:req}) and (\ref{eq:Pieq}) are order $1/\epsilon^2$ and $1/\epsilon$ respectively.
Correspondingly, these equations are unaltered in the $\epsilon \to 0$ limit.   The scaling relation (\ref{eq:Piscaling}) and
the definition of $\Pi_a$ in (\ref{eq:Pidef}) imply that at leading order in $\epsilon$
the metric component $F^a$ satisfies
\begin{equation}
\label{eq:Feqmleading}
\partial_\lambda F^a = 0.
\end{equation}
With the boundary condition (\ref{eq:FMBC}), this equation has the solution
\begin{equation}
\label{eq:Fsol}
F^a = \Omega \delta^{a \phi}.
\end{equation}
The solution (\ref{eq:Fsol}) also satisfies the radial constraint
equation (\ref{eq:Fdoteq}) at leading order in $\epsilon$.

Employing (\ref{eq:Fsol}), at leading order in $\epsilon$ the remaining Einstein equations (\ref{eq:rdoteq}), (\ref{eq:hdoteq}), (\ref{eq:Aeq}) and (\ref{eq:rddoteq}), as well as the scalar equation of motion (\ref{eq:scalareq}), respectively reduce to 
\begin{subequations}
\label{eq:einsteinscalar}
\begin{align}
\label{eq:rdoteq2}
0 &= (r D_+ r)', &
\\ \label{eq:hdoteq2}
0 &= \ts \left [ \delta^{c}_{a} \delta^{d}_b - \frac{1}{2}G^{cd} G_{ab}\right ] (r D_+ h_{cd})'+D_+ r h'_{ab}, &
\\ \label{eq:Aeq2}
0 &= \ts A'' + \ts \frac{1}{4}  G'^{ab} D_+ G_{ab}  - \frac{4 r'}{r^2} D_+ r  + 
8 \pi \Psi' D_+ \Psi, &
\\ \label{eq:rddot2}
0 & \ts = D_+^2 r - A' D_+ r  +\frac{r^5}{8}  D_+ h^{ab} D_+ h_{ab}
+ 4 \pi r (D_+ \Psi)^2,&
\\ \label{eq:scalareq2}
0 &= (r D_+ \Psi)' + \Psi' D_+ r. &
\end{align}
\end{subequations}
No $\theta$ derivatives appear in the equations of motion (\ref{eq:einsteinscalar})
and $\phi$ derivatives only appear in the combination $\partial_v + \Omega \partial_\phi$.  It follows that the $\phi$ dependence always comes in the combination $ \phi - \int^v d \tilde v \Omega(\tilde v) $.  Hence the equations of motion (\ref{eq:einsteinscalar}) constitute a system of 1+1D PDEs at each set of angles.  Of course this can be made more explicit by working in a rotating reference frame by setting $\Omega = 0$.

Eq.~(\ref{eq:rdoteq2}) can be integrated to yield
\begin{equation}
\label{eq:rdot}
D_+ r = -\frac{\zeta}{r}.
\end{equation}
The constant of integration $\zeta$ can be determined from the radial constraint equation (\ref{eq:rddot2}).  Since the total time
derivative of any quantity $f$ on $\M$ is $df/dv = D_+ f|_{\lambda = 0}$,
on $\M$ Eq.~(\ref{eq:rddot2}) can be rewritten 
\begin{equation}
\label{eq:constraintlove}
\frac{d}{dv} D_+ r + \kappa D_+ r =  -\frac{r^5}{8}  D_+ h^{ab} D_+ h_{ab}
- 4 \pi r (D_+ \Psi)^2,
\end{equation}
where we have employed (\ref{eq:kappaKerr}) to eliminate $A'$ in favor of the surface gravity $\kappa$. How does this equation behave as $v \to \infty$?  Under the assumption that gravitational wave tails decay faster than scalar tails (meaning $q > p$), the first term of the r{.}h{.}s{.} can be neglected. 
Likewise, since $v$ derivatives on $\M$ are small compared to $\kappa$, the first term on the l{.}h{.}s{.} can also be neglected.  Therefore, at late times Eq.~(\ref{eq:constraintlove}) becomes an algebraic equation for $D_+r$ on $\M$.  Using (\ref{eq:rdot}) we therefore find   
\begin{equation}
\label{eq:zetageneral}
\zeta \sim (D_+ \Psi)^2|_{\lambda = 0}.
\end{equation}
Using Price's Law (\ref{eq:scalarinflux}), this becomes
\begin{equation}
\label{eq:zetaRN}
 \zeta \sim +1/v^{2p}.
\end{equation}

Eqs.~(\ref{eq:rdot}) and (\ref{eq:zetaRN}) can be employed to study the causal structure of the spacetime enclosed by $\M$.
Consider an arbitrary null curve with velocity $\dot x^\mu = \{1,\dot \lambda,\dot x^a\}$ with $\dot f \equiv \partial_v f$ for any $f$.  
From the chain rule the radial velocity reads
\begin{equation}
\label{eq:radialvel}
\dot r = \partial_v r + \D_a r  \dot x^a + r' \dot \lambda.
\end{equation}
The null condition $g_{\mu \nu} \dot x^\mu \dot x^\nu = 0$
requires
\begin{equation}
\label{eq:nullcondition}
\dot \lambda =  A - \frac{1}{2}G_{ab}(F^a - \dot x^a)(F^b - \dot x^b).
\end{equation}
Substituting (\ref{eq:nullcondition}) into (\ref{eq:radialvel}) yields
\begin{equation}
\dot r = D_+ r - (F^a - \dot x^a) \D_a r - \frac{1}{2} r' G_{ab}(F^a - \dot x^a)(F^b - \dot x^b).
\end{equation}
The radial velocity is maximized when $F^a - \dot x^a = - \frac{1}{r'} \D^a r$.  It therefore follows that the maximum radial velocity of null curves is 
\begin{equation}
\label{eq:maxdotr1}
\max(\dot r) = D_+ r + \frac{1}{2 r'} (\D r)^2.
\end{equation}
The last term on the r.h.s. of this equations is order $\epsilon$ (and also turns out to vanish at $r = 0$ at late times).  Employing Eqs.~(\ref{eq:rdot}) and (\ref{eq:zetaRN}), at leading order 
in $\epsilon$ we therefore obtain
\begin{equation}
\label{eq:outgoinggeos}
\max(\dot r) \sim - \frac{v^{-2p}}{r}.
\end{equation}

Eq.~(\ref{eq:outgoinggeos}) means that within the domain of validity of the derivative expansion, 
all light rays propagate to smaller $r$.  
Eq.~(\ref{eq:outgoinggeos}) can be integrated in closed form to yield
\begin{equation}
\label{eq:geoeq}
r^2 \sim v^{1- 2 p} + {\rm const.}
\end{equation}
Clearly \textit{all} null curves at  $r < r_{\rm c}$ must end at $r = 0$ in a finite time,
where the \textit{critical radius} $r_{\rm c}$ reads
\begin{equation}
\label{eq:logkicksin}
r_{\rm c}  \sim v^{1/2 - p}.
\end{equation}
Light rays with $r > r_{\rm c}$ can end on the CH at finite values of $r$.  This behavior is
consistent with the Penrose diagrams shown in Figs.\ref{fig:PenroseDiagram} and \ref{fig:CompDomain}.

The remaining equations of motion (\ref{eq:hdoteq2}), (\ref{eq:Aeq2}) and (\ref{eq:scalareq2}) cannot
be solved analytically without further approximations.  In Sec.~\ref{sec:nearM} we shall solve these equations away from $r = 0$, where infalling radiation can treated perturbatively, and establish the self-consistency condition that $\lambda$ derivatives blow up like $e^{\kappa v}$.  In Sec.~\ref{sec:nearorigin}
we shall show that at $r \lesssim r_{\rm c}$, $\lambda$ derivatives
blow up like $r^{ -\alpha v} e^{\kappa v}$ where $\alpha > 0$ is some constant.  

\subsection{Behavior away from $r = 0$}
\label{sec:nearM}

In this section we solve Eqs.~(\ref{eq:einsteinscalar})
away from $r = 0$. Since infalling radiation decays as $v \to \infty$,
we can neglect its effects away from $r = 0$ late times.  This is tantamount to imposing the boundary conditions 
$D_+ r = D_+ h_{ab} = D_+ \Psi = 0$ on $\M$.  In this case Eqs.~(\ref{eq:rdoteq2}), (\ref{eq:hdoteq2}), (\ref{eq:Aeq2}) and (\ref{eq:scalareq2}) reduce to
\begin{align}
\label{eq:optics}
&D_+ r = 0, &
&D_+ \Psi = 0,&&
D_+ h_{ab} = 0,&& A'' = 0. 
\end{align}
The first three equations here simply state that excitations in $r$, $\Psi$ and $h_{ab}$ are transported
along null curves tangent to $D_+$.
Using the boundary conditions (\ref{eq:AKerr}) and (\ref{eq:kappaKerr}), the solutions to (\ref{eq:optics}) read
\begin{subequations}
\label{eq:outfluxsolutions}
\begin{align}
h_{ab} &= \mathcal H_{ab}(e^{\kappa v}\lambda,\theta,\varphi), \\
\Psi &= \chi(e^{\kappa v}\lambda,\theta,\varphi), \\
r &= \mathfrak R(e^{\kappa v}\lambda,\theta,\varphi), \\ \label{eq:AsolnearM}
A & = -\kappa \lambda,
\end{align}
\end{subequations}
where $\varphi \equiv \phi- \int^v d\tilde v \Omega(\tilde v)$.
Note $\mathfrak R$ is determined by $\chi$ and $\mathcal H_{ab}$ via Eq.~(\ref{eq:req}).
The function $\chi$ encodes an outgoing flux
of scalar radiation inside the black hole while $\mathcal H_{ab}$ encodes an outgoing flux of gravitational radiation.  
The outgoing radiation is not unique and reflects the fact that the 
geometry inside the black hole depends on past evolution, or in our framework, on initial data at $v = v_o$. 
Note that the only non-trivial boundary condition we imposed on $\M$, Eq.~(\ref{eq:kappaKerr}), is in fact valid away from $\M$.  It follows that our analysis is insensitive to the precise choice of $\M$.
We could have just as well chosen $\M$
to asymptote to some finite $r < r_-$ as $v \to \infty$.

The solutions (\ref{eq:outfluxsolutions}) are consistent with the scaling
hypothesis (\ref{eq:scalings}) with bookkeeping parameter $\epsilon \equiv e^{-\kappa v}$, meaning 
\begin{equation}
\partial_\lambda \sim e^{\kappa v}.
\end{equation}
From Raychaudhuri's equation (\ref{eq:req}) we see that $r'$ can only increase as $\lambda$,
or equivalently $r$, decreases. This means that $r= 0$ lies at 
$\lambda = O(e^{-\kappa v})$.  Correspondingly, the solution (\ref{eq:AsolnearM}) implies $A = O(e^{-\kappa v})$, which is consistent with the scaling relation
(\ref{eq:Ascalinghyp}) with $\epsilon = e^{-\kappa v}$. 

As stated in the Introduction, the physical origin of large $\lambda$ derivatives lies in the fact that from the perspective of infalling observers, outgoing radiation inside  $r_-$ appears blue shifted by a factor of $e^{\kappa v}$.   This blue shift results in dynamics at different angles becoming causally disconnected from each other.  Consequently, at $r < r_-$ the 
equations of motion at different angles decouple, resulting in a system of 1+1D PDEs at each set of angles, just as we have found.

\subsection{Behavior at small $r$}
\label{sec:nearorigin}

The analysis in the preceding section neglected the effects of 
infalling radiation.  However, the effects of infalling radiation
cannot be neglected at small $r$. Indeed, $D_+ r \sim -\frac{v^{-2 p}}{r}$ diverges at $r = 0$ with an amplitude determined by infalling radiation.
In this section we show that at small $r$, infalling radiation results in derivatives w{.}r{.}t{.}~$\lambda$ blowing up like $r^{-\alpha v} e^{\kappa v}$, where $\alpha$ is a positive constant. 

Following \cite{CNC}, to study the behavior of the fields at small $r$ we have found
it convenient to change radial coordinates from $\lambda$ to $r$.
In the $\{v,r,\theta,\phi\}$ coordinate system Eq.~(\ref{eq:Dplus}) becomes 
\begin{align}
\label{eq:dplusrsystem}
D_+ &= \partial_v + (D_+ r) \partial_r + F^a \partial_a,
\\ \nonumber
&= \ts \partial_v - \frac{\zeta}{r} \partial_r + \Omega \partial_\phi,
\end{align}
where in the last line we used (\ref{eq:rdot}) and (\ref{eq:Fsol}).
The tensor and scalar wave equations (\ref{eq:hdoteq2}) and (\ref{eq:scalareq2}) become
\begin{subequations}
	\label{eq:scalartensor}
\begin{align}
\label{eq:scalarwaver}
0 &= \ts \partial_r (r D_+ \Psi)- \frac{\zeta}{r} \partial_r \Psi,
\\ \label{eq:tensorwaver}
0 &= \ts \left [ \delta^{c}_{a} \delta^{d}_b - \frac{1}{2}G^{cd} G_{ab}\right ] \partial_r (r D_+ h_{cd})-  \frac{\zeta}{r} \partial_r  h_{ab}, 
\end{align}
\end{subequations}
where we have again used (\ref{eq:rdot}) to eliminate $D_+ r$.
We therefore reach the conclusion that in the $\{v,r,\theta,\phi\}$ coordinate system,
the scalar and tensor modes satisfy 
decoupled PDEs.  Remarkably, the scalar equation of motion is linear.

The equations of motion (\ref{eq:scalartensor}) can be solved analytically near $r = 0$.  Approximating $D_+ \approx - \frac{\zeta}{r} \partial_r$, the scalar equation of motion (\ref{eq:scalarwaver}) has the solution 
\begin{align}
\label{eq:scalarnearorigin}
\Psi = f \log m r.
\end{align}
The constants of integration $f$ and $m$ encode the flux of ingoing and outgoing radiation.  Likewise, with the approximation $D_+ \approx - \frac{\zeta}{r} \partial_r$ the tensor equation of motion (\ref{eq:tensorwaver}) has the solution
\begin{align}
\label{eq:tensornearorigin}
& h_{\theta \phi}  = {\beta} \sin \theta \sinh\left ( {{\ts  \gamma}} \log \mu r \right) , &&
\frac{h_{\theta \theta}}{h_{\phi \phi}} = \csc^2 \theta \frac{1 + \sqrt{1 - \beta^2} \tanh \left ({{\ts  \gamma}} \log \mu r \right )}{1 - \sqrt{1 - \beta^2} \tanh \left ({{\ts  \gamma}} \log \mu r \right )} e^{2 \nu}.
\end{align}
Together with the normalization condition (\ref{eq:hnorm}), these equations determine $h_{ab}$ near $r = 0$.
The four constant of integration $\beta, \mu, \nu$ and $\gamma$ encode the amplitudes of infalling and outgoing radiation for the two gravitational modes. 

The equations of motion (\ref{eq:scalartensor}) imply the ``energy densities"
\begin{align}
&\mathcal E \equiv r (\partial_r \Psi)^2, &&\rho \equiv  r \, {\rm tr} \left (h^{-1} \partial_r h  h^{-1} \partial_r h \right ),
\end{align}
satisfy the conservation laws 
\begin{align}
\label{eq:scalarcons}
(\partial_v + \Omega \partial_\phi) \mathcal E + \partial_r \mathcal S &= 0,
\\ \label{eq:tensorcons}
(\partial_v + \Omega \partial_\phi) \rho + \partial_r J &= 0,
\end{align}
where the fluxes $\mathcal S$ and $J$ read
\begin{align}
\label{eq:scalarflux}
\mathcal S &\equiv \ts \frac{r^2 }{\zeta}  (D_+ \Psi)^2 - \zeta \, (\partial_r \Psi)^2,
\\
\label{eq:tensorflux}
 J &\equiv \ts \frac{ r^2}{\zeta} {\rm tr} \left (h^{-1} D_+ h  h^{-1} D_+ h \right ) - \zeta {\rm tr} \left (h^{-1} \partial_r h  h^{-1} \partial_r h \right ).
\end{align}
The behavior of $\mathcal E$ and $\rho$ near $r = 0$ will play a central role in our analysis.  Eqs.~(\ref{eq:scalarnearorigin}) and (\ref{eq:tensornearorigin}) imply that near $r = 0$ the energy densities diverge like  
\begin{subequations}
	\label{eq:energyscalings}
\begin{align}
\label{eq:energyscalingsscalar}
\mathcal E &\sim \frac{ f^2}{r}, \\
\label{eq:energyscalingstensor}
\rho &\sim \frac{2 \gamma^2}{r}.
\end{align} 
\end{subequations}

To proceed further we must specify boundary conditions on $\M$.  
The amplitude of infalling 
radiation is fixed by specifying $D_+ \Psi$ and $D_+ h_{ab}$ while the amplitude of outgoing radiation is fixed by specifying $\partial_r \Psi$ and $\partial_r h_{ab}$. Note that the contributions to the fluxes $\mathcal S$ and $J$ from outgoing radiation is suppressed by $\zeta$, which according to (\ref{eq:zetaRN}) vanishes at late times.  The fact that outgoing modes 
don't contribute to the fluxes at $\M$ is simply a consequence of the fact that
outgoing light rays have vanishing radial velocities at late times.
Indeed, the outgoing characteristic curves of the differential operators in  (\ref{eq:scalartensor}) satisfy $dr/dv \sim -v^{-2 p}$.

We employ Price's law (\ref{eq:PriceLaw0}) for the amplitude of infalling radiation. 
Using (\ref{eq:zetaRN}) we then see that as $v\to \infty$ the fluxes through $\M$ read
\begin{align}
\label{eq:fluxthroughM}
&\mathcal S|_{\M} \sim \frac{v^{-2 p}}{\zeta} \sim  1, &&
J|_{\M} \sim \frac{v^{-2 q}}{\zeta} \sim v^{2(p -q)}. 
\end{align}
The scalar flux is constant in time.  Moreover, owing to the fact that at late times the explicit time dependence in the scalar equation of motion (\ref{eq:scalarwaver}) is arbitrarily slowly varying, the transport of scalar energy must approach a steady-state.
Near $r = 0$ this means $\partial_v^2 f^2 = 0$.  It follows that as $v \to \infty$ we must have
\begin{equation}
\label{eq:scalarenergyscaling}
f \sim \sqrt{v}.
\end{equation}
In contrast, due to the assumption $q > p$, the tensor energy 
flux vanishes at late times. It follows that $\gamma^2 \ll f^2$ as $v\to \infty$.
In other words, at small $r$ the tensor energy density is negligible compared to the scalar energy density.

Let us briefly address the domain of validity of the 
approximate scalar solution (\ref{eq:scalarnearorigin}).  Using (\ref{eq:scalarenergyscaling}) and (\ref{eq:scalarnearorigin}) we see that 
$\partial_v \Psi \sim \frac{\zeta}{r} \partial_r \Psi$ when 
$r \sim r_{\rm c}$.
Therefore, neglecting $v$ derivatives in $D_+$ becomes valid when $r \lesssim r_{\rm c}$, meaning the expression (\ref{eq:energyscalingsscalar}) becomes valid when $r \lesssim  r_{\rm c}$.  Evidently, the tiny decaying influx of scalar radiation through $\M$ results in a cloud of scalar radiation forming at $r \lesssim r_{\rm c}$. The energy density of the cloud grows like $\mathcal E \sim v/r$, becoming increasingly singular as $v \to \infty$.

Let us now return to using the affine parameter $\lambda$ 
as a radial coordinate and investigate the consequences of 
of the diverging energies (\ref{eq:energyscalings}) on the behavior of $r'$ as $r \to 0$.  First, note Raychaudhuri's equation
(\ref{eq:req}) can be written
\begin{equation}
\label{eq:reqnew}
\ts r'' + r'^2 \left [\frac{1}{8} \rho + 4 \pi \mathcal E \right ] = 0.
\end{equation}
Using Eqs.~(\ref{eq:energyscalings}), Eq.~(\ref{eq:reqnew}) may be integrated near $r = 0$ to yield 
\begin{equation}
\label{eq:rprimediv}
r' \sim r^{-z} C,
\end{equation}
where $C$ is a constant of integration and 
\begin{equation}
\label{eq:zdef}
z\equiv 4 \pi f^2 + \frac{\gamma^2}{4}.
\end{equation}
Recall that near $\M$ we have $r' \sim e^{\kappa v}$ and that 
$r'$ can only increase as $r$ decreases.  It follows that $C \sim e^{\kappa v}$.
Using Eq.~(\ref{eq:scalarenergyscaling}), we then see that as $v \to \infty$
\begin{equation}
\label{eq:rscalingenhanced}
r' \sim r^{-\alpha v} e^{\kappa v},
\end{equation}
where $\alpha > 0$ is a constant. 
Since the logarithmic divergence in (\ref{eq:scalarnearorigin})
becomes dominant when $r \lesssim r_{\rm c}$, the $r^{-\alpha v}$ enhancement also kicks in when $r \lesssim r_{\rm c}$.  
It follows from the chain rule that for $r \lesssim r_{\rm c}$, derivative w{.}r{.}t{.}~$\lambda$ blow up like
\begin{equation}
\partial_\lambda \sim r^{-\alpha v} e^{\kappa v}.
\end{equation}
Hence the diverging energy densities $\mathcal E$ and $\rho$ lead to 
diverging $\lambda$ derivatives at $r = 0$.
Evidently, the increasingly singular cloud of
scalar radiation \textit{focuses} infalling light rays, 
hastening the affine parameter interval between $r = r_{\rm c}$ and $r = 0$. 

\subsection{Domain of dependence and domain of validity of initial value problem}
\label{sec:consistancy}

At small $r$ the approximate equations of motion (\ref{eq:einsteinscalar}) can break down.  Why?
In deriving the approximate equations of motion we have neglected terms which diverge like 
$1/r^n$ near $r = 0$ for some set of fixed powers $n$.  For example, the Ricci scalar $\mathcal R$ of the angular metric $G_{ab}$ diverges as $r\to 0$. In deriving Eqs.~(\ref{eq:rdoteq2}) and (\ref{eq:Aeq2}) from Eqs.~(\ref{eq:rdoteq}) and (\ref{eq:Aeq}), we have neglected $\mathcal R$.
If initial data at $v = v_o$ is non-singular, with $\lambda$ derivatives of order $e^{\kappa v_o}$ but finite at $r = 0$, then the approximate equation of motion break down beyond $\log \frac{1}{r} \sim v_o$, where the neglected $1/r^n$ power law divergences in Einstein's equations become comparable to $\lambda$ derivatives.  

However, even with non-singular initial data, the analysis in the preceding section demonstrates infalling radiation results in a cloud of scalar radiation forming at $r \lesssim r_{\rm c}$, which grows unboundedly large as $v \to \infty$.  The cloud enhances the 
focusing of infalling geodesics and results in $\lambda$ derivatives
diverging like $r^{-\alpha v} e^{\kappa v}$ inside $r_{\rm c}$.  This means that at late enough times, $\lambda$ derivatives are always larger than the  neglected $1/r^n$ divergent terms and the the approximate equations of motion (\ref{eq:einsteinscalar}) are valid all the way to $r = 0$.

Within the domain of validity of the approximate equations of motion (\ref{eq:einsteinscalar}), what initial data at $v = v_o$ can we propagate forward in $v$?  
Answering this equation is tantamount to specifying the outgoing
null surface $\mathcal F$ shown in Fig.~\ref{fig:CompDomain}.   Let us restrict our attention to outgoing null surfaces governed by evolution equation (\ref{eq:geoeq}).  We parameterized initial data for $\mathcal F$ via
\begin{equation}
r(v_o)^2 = r_{\rm c}(v_o)^2 (1 - \delta),
\end{equation}
with $0<\delta < 1$.  The evolution equation (\ref{eq:geoeq}) then yields
\begin{equation}
\label{eq:Fdef}
r^2 \sim v^{1 - 2p} - v_o^{1 - 2p} \delta.
\end{equation}
On $\mathcal F$ we then compare $1/r^n$ to $\lambda$ derivatives.

Consider first the limit $\delta \to 0$.  In this case $\mathcal F$ coincides with the critical radius $r_{\rm c}\sim v^{1/2-p}$ and intersects $r = 0$ at $v = \infty$.  On $\mathcal F$ the power law divergences scale like $1/r^n \sim v^{n (p-1/2)}$.  In contrast, $\lambda$ derivatives scale like $e^{\kappa v}$ and hence are always parametrically larger than the $1/r^n$ divergences, irrespective of the 
$r^{-\alpha v}$ late-time enhancement. 
Next, suppose $\delta$ is arbitrarily small but finite.
In this case Eq.~(\ref{eq:Fdef}) implies that $\mathcal F$ intersects 
$r = 0$ at $v = v_*$ where $v_{*}= v_o/ \delta^{1/(2 p - 1)}$.  This means the $1/r^n$ terms diverge on $\mathcal F$ at time $v = v_* \gg v_o$. Do $\lambda$ derivatives diverge faster than the neglected $1/r^n$ terms?  The answer is clearly yes due to the $r^{-\alpha v}$ late-time enhancement: the exponent $\alpha v_*$ can be made arbitrarily large by taking $\delta$ smaller whereas the exponent $n$ is fixed.  Hence $\lambda$ are exponentially larger than the neglected $1/r^n$ divergent terms.  This means that, as depicted in Fig.~\ref{fig:CompDomain}, the domain of dependence and validity of our initial value problem always contains $r = 0$ at late enough times. 
We may therefore use the approximate equations of motion to study the nature of the geometry at $r = 0$ at late enough times.

\section{Singular structure of the spacetime}
\label{sec:singularstruct}

The divergence of $\lambda$ derivatives at $v = \infty$ and $r = 0$
indicates the presence of singularities. Consider for simplicity the Ricci scalar,
\begin{equation}
R = 16 \pi \Psi' D_+ \Psi + 8 \pi (\D \Psi)^2.
\end{equation}
At $r \gtrsim r_{\rm c}$ derivatives w{.}r{.}t{.} $\lambda$ scale like $e^{\kappa v}$.  Hence at $r \gtrsim r_{\rm c}$ the Ricci scalar blows up like 
\begin{align}
\label{eq:outsiderc}
	R \sim e^{\kappa v}, 
\end{align}
indicating a null singularity at the geometry's CH.  Likewise,
at $r \lesssim r_{\rm c}$ derivatives w{.}r{.}t{.} $\lambda$ scale like $r^{-\alpha v}e^{\kappa v}$.  Hence at $r \lesssim r_{\rm c}$ we have 
\begin{align}
\label{eq:insiderc}
R \sim r^{-\alpha v} e^{\kappa v}, 
\end{align}
indicating a central singularity at $ r= 0$. Note for the 
Kretschmann  scalar $K \equiv R^{\mu \nu \alpha \beta}R_{\mu \nu \alpha \beta}$, the analogous scaling relations are just the square of Eqs.~(\ref{eq:outsiderc}) and (\ref{eq:insiderc}).

Since all null curves at $r < r_{\rm c}$
end at $r = 0$ at finite $v$, the singularity at $r = 0$ must be spacelike.  This, combined with the singular CH, means the singular structure of the spacetime is that depicted in Fig.~(\ref{fig:CompDomain}).
It is noteworthy that the magnitude of the singularity at $r = 0$ increases as a function of time $v$.  This is due to the 
cloud of scalar radiation forming at $r \lesssim r_{\rm c}$, which grows unboundedly large in magnitude as $v \to \infty$
due to being sourced by infalling tails.    

The physical origin of the singularity at $v = \infty$ to understand \cite{Penrose:1968ar,Poisson:1989zz,PhysRevD.41.1796,CNC}.  Consider two \textit{crossing fluxes} of scalar radiation of wavelength $L_+$ and $L_-$. The crossing fluxes result in $R \sim \frac{1}{(L_+ L_-)}$.  In the limit where either $L_\pm  \to 0$, the curvature scalar diverges.  
Inside the black hole Price's law (\ref{eq:PriceLaw0}) provide an flux of 
infalling radiation.  Likewise, the interior geometry 
of the black hole is filled with outgoing radiation, which need not fall into $r = 0$.
From the perspective of infalling observers, the outgoing radiation appears blue shifted by $e^{\kappa v}$ (which manifests itself in our coordinate system as $\lambda$ derivatives growing like $e^{\kappa v}$) 
\footnote{From the perspective of \textit{outgoing} observers, ingoing radiation appears blue shifted by $e^{\kappa v}$.}.
Correspondingly, the crossing fluxes must result in the exponentially diverging curvatures, just as seen in Eqs.~(\ref{eq:outsiderc}) and (\ref{eq:insiderc}).

\section{Discussion}
\label{sec:discussion}

While the geometry inside black holes depends on initial conditions,
our analysis demonstrates that near $r = 0$ --- specifically at $r \lesssim r_{\rm c}$ --- the geometry 
enjoys universal features.  The universal features arise not due to initial data per se, but rather from the universal nature of decaying influxes of radiation 
into the black hole.  The decaying fluxes result in a cloud of scalar radiation forming at $r \lesssim r_{\rm c}$, with the scalar field $\Psi$ growing like $\Psi \sim \sqrt{v} \log r$.  The growing scalar field results in the curvature diverging like $r^{-\alpha v}$, where $\alpha > 0$ is a constant.  

It is noteworthy that the growth of the scalar cloud doesn't depend on the power $p$ associated with Price's Law (\ref{eq:scalarinflux}).  
This can be traced back to the fact that the scalar energy flux $\mathcal S$ through $\M$ is $v$-independent. 
In particular, the $v$-dependence of $\mathcal S$ coming from $D_+ \Psi$ in Eq.~(\ref{eq:scalarflux})
is canceled by that of $\zeta$.  In fact this happens irrespective of if the influx decays with a power of $v$.  One merely needs the scalar influx to decay slower than both the gravitational influx and $e^{-\kappa v}$.  With these two assumptions, Eq.~(\ref{eq:zetageneral}) is in fact still valid, and the scalar energy flux $\mathcal S$ through $\M$ is still $v$-independent.  
This means that our analysis can readily be generalized to de Sitter spacetime, where influxes decay exponentially in $v$ instead of with power laws.

\begin{figure}[ht]
	\includegraphics[trim= 0 0 0 0 ,clip,scale=0.35]{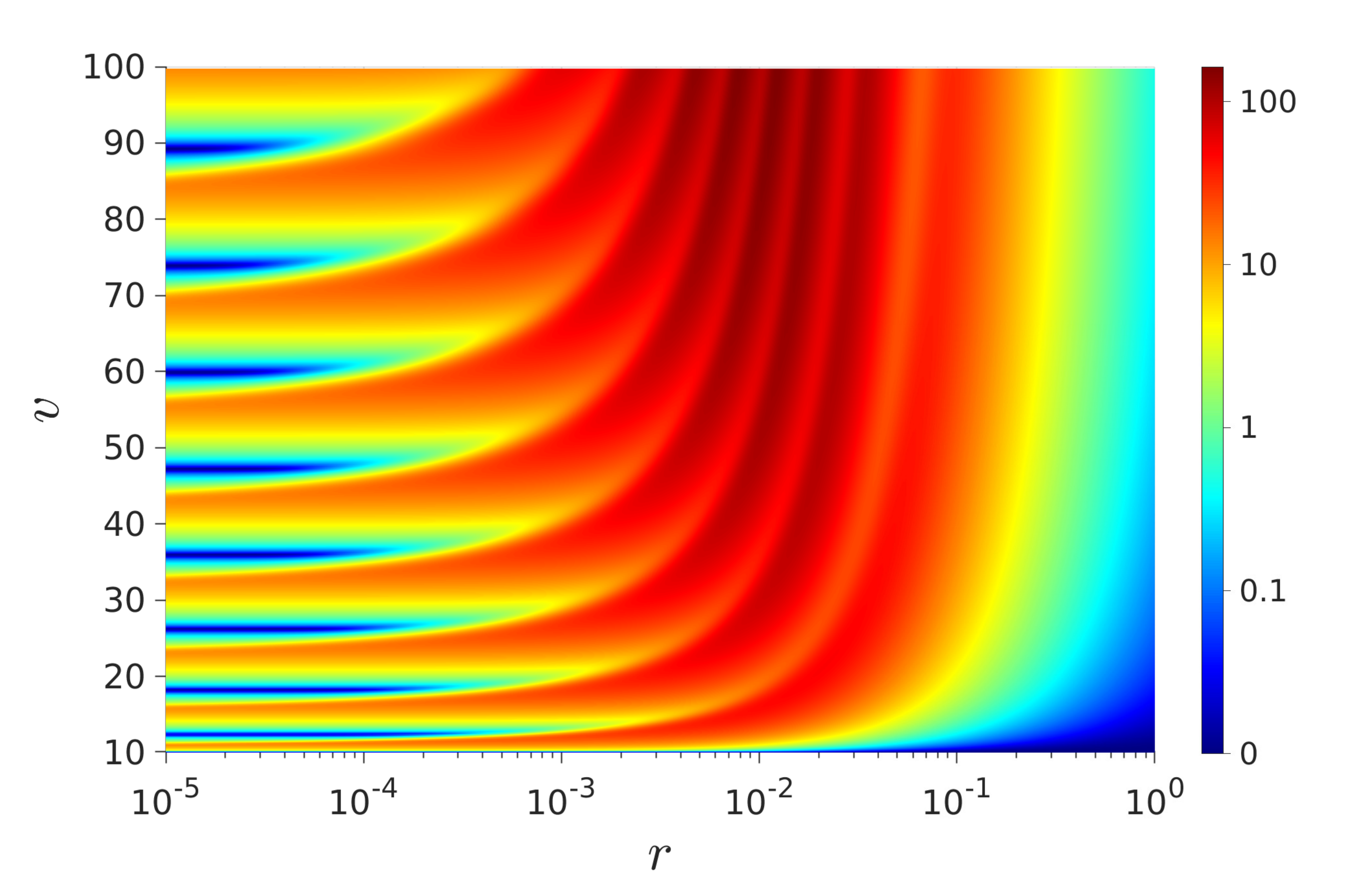}
	\caption{
		$r \rho$ for a numerically generated solution to the tensor equations of motion (\ref{eq:tensorwaver}).  $\rho$ periodically vanishes at small $r$.
	}
	\label{fig:tensormodes}
\end{figure}

As stated in the Introduction, without a scalar field, the results of Belinsky, Khalatnikov, and Lifshitz suggest that any spacelike singularity at $r = 0$ must be oscillatory in nature \cite{Belinsky:1970ew,doi:10.1063/1.3382327}.  This contrasts with the monotonic singularity found in this paper.  It is therefore natural to ask how much of our analysis carries over to vacuum solutions.  With the scalar turned off, do the approximate equations of motion (\ref{eq:einsteinscalar}) still hold all the way to 
$r = 0$? The answer appears to be no.  

Let us consider then the vacuum equations.  In vacuum 
Eq.~(\ref{eq:zetaRN}) is replaced by 
\begin{equation}
\label{eq:zetagrav}
\zeta \sim + 1/v^{2 q}.
\end{equation}
This follows from integrating Eq.~(\ref{eq:constraintlove}) with Price's law (\ref{eq:tensorinflux}) and vanishing scalar.  The tensor energy flux (\ref{eq:tensorflux}) through $\M$ therefore scales like
\begin{equation}
\label{eq:tensorflux1}
J|_{\M} \sim \frac{v^{-2 q}}{\zeta} \sim 1.
\end{equation}
Naively, Eq.~(\ref{eq:tensorflux1}) suggests that the tensor energy density $\rho$ grows linearly in $v$ near $r = 0$, just like the scalar energy $\mathcal E$ did.
However, unlike the scalar field, the tensor field satisfies a nonlinear equation of motion, Eq.~(\ref{eq:tensorwaver}). Due to nonlinear interactions, the flux $J$ need not be constant in time near $r = 0$ and can in fact oscillate. 
To demonstrate this, in Fig.~\ref{fig:tensormodes} we plot the tensor energy density $\rho$ for a numerically generated solution to (\ref{eq:tensorwaver}).  
We see from the figure that $\rho$ develops oscillations in $v$ and $r$ and occasionally vanishes at small radii.  This stands in stark contrast to the scalar energy density, which grows like $\mathcal E \sim v/r$.  From Raychaudhuri's equation, Eq.~(\ref{eq:reqnew}), the vanishing of $\rho$ means that there is no enhancement of $\lambda$ derivatives at small radii.  Correspondingly, when $\rho$ vanishes, the approximate equations of motion break down beyond $\log \frac{1}{r} \sim v$, where neglected power law divergences in Einstein's equations become comparable to $\lambda$ derivatives.
It remains to be seen how universal the behavior seen in Fig.~\ref{fig:tensormodes} is.
We leave a detailed study of singularities in vacuum for future work.

\section{Acknowledgments}%
This work is supported by the Black Hole Initiative at Harvard University, 
which is funded by a grant from the John Templeton Foundation. We thank Amos Ori, Jordan Keller and Ramesh Narayan for many helpful conversations during the preparation of this paper.

\section*{References}
\bibliography{refs}%
\bibliographystyle{utphys}

\end{document}